\documentstyle[prd,aps]{revtex}

\begin{document}

\title{Quantum processes \\ beyond the Aharonov--Bohm Effect}

\author{J\"urgen Audretsch$^{\dagger,}$\cite{bylineaud} \and Vladimir D.~Skarzhinsky$^{\ddagger,}$\cite{bylineskarzh}}

\address{$^\dagger$Fakult\"at f\"ur Physik der Universit\"at Konstanz\\
Postfach 5560, M 674, D-78434 Konstanz, Germany \and \\ $^\ddagger$P.~N.~Lebedev Physical Institute\\
Leninsky prospect 53, Moscow 117924, Russia}

\date{Received 30 May 1997}

\maketitle

\vskip1cm
\begin{center}
{\small\sl This article is dedicated to the memory of Prof. Asim O. Barut \\ with whom we had many discussions about the Aharonov-Bohm effect.}
\end{center}

\begin{abstract}

\noindent
We consider QED - processes in the presence of an infinitely thin and infinitely long straight string with a magnetic flux inside it. The  bremsstrahlung from an electron passing by the magnetic string and the electron-positron pair production by a single photon are reviewed. Based on the exact electron and positron solutions of the Dirac equation in the external Aharonov--Bohm potential we present matrix elements for these processes. The dependence of the resulting cross sections on energies, directions and polarizations of the involved particles is discussed for low energies.

\end{abstract}


\vspace{0.5cm}


\section{Introduction}

The influence of magnetic fluxes on charged quantum systems -- the
Aharonov--Bohm (AB) effect \cite{Aharonov59} -- was investigated theoretically by many authors, and it was observed in numerous electron interference experiments. Now it is well established and generally recognized effect. For a comprehensive review see \cite{Olariu85,Peshkin89}.

However, the traditional applications of the AB effect were up to now the treatment of electron scattering and interference as well as of bound states. 
For a deeper understanding of the AB effect and in order to find additional experimental verifications, it is therefore highly desirable to study further applications of the effect. With regard to this a detailed investigation of QED processes in the presence of the magnetic string has recently been carried out \cite{Audretsch96',Audretsch96''}. Among these processes the bremsstrahlung from the Dirac electron passing by the magnetic string \cite{Audretsch96'} seems to be the most obvious one since it is related directly as an accompanying effect to the AB scattering. It was discussed at first for non-relativistic particles in the dipole approximation in \cite{Sereb88}, where the energy and angular distributions were calculated. The polarization properties of the bremsstrahlung for relativistic spinless particles were studied in \cite{Gal'tsov90}. The electron-positron pair production by a single photon in the AB field has been treated as well \cite{Audretsch96''}. It looks more mysteriously since photons do not interact with the magnetic string directly. It goes back to the created particles which are affected by the magnetic flux.

These quantum processes, with only a single particle in the respective initial states, are forbidden by the energy-momentum conservation law unless external fields are present for the necessary energy-momentum transfer. This may be for example a  Coulomb field \cite{Akhiezer65} or an uniform magnetic field \cite{Klepikov54,Ginzburg65}. In these two cases external local forces are present which influence the motion of created charged particles. In contrast with this, in the AB potential the processes happen for a global, topological reason. In this regard quantum processes near magnetic strings bear some resemblance with processes near cosmic strings \cite{Skarzhinsky94,Audretsch94}.

In the present paper we summarize the main results obtained recently for QED processes in the AB potential and discuss their peculiar features. We hope this will stimulate new interest in the theoretical discussion and the experimental investigation of the AB effect.

We consider the simple idealized case of an infinitely thin, infinitely long straight magnetic string lying along the $z$ axis. It is described by the singular AB--potential. This leads to the non-selfadjointness of the Hamilton operator \cite{Hagen90,Gerbert89} and raises the question how to describe wave functions near the string correctly. We follow the papers \cite{Hagen90,Audretsch95} in this point. In section 2 we describe exact solutions of the Dirac equation in the AB potential in terms of the partial waves as well as the scattering wave functions. It is worth to notice that, due to the interaction between spin and magnetic field, the wave functions for spin particles do not vanish on the magnetic string, and thus the AB effect loses its nonlocal status.

In section 3 we represent the matrix elements for the bremsstrahlung emitted by an electron passing by the magnetic string, and for the pair production by a single photon. The behavior of the differential and total cross sections at low energies and particular features of these processes for the Dirac particles are described in section 4. We conclude in section 5 with a discussion of the obtained results.
\medskip

We use units such that $\hbar=c=1$ and take $e<0$ for the electron
charge.


\section{The electron and positron solutions to the Dirac equation in
the Aharonov--Bohm potential}

The AB vector potential for the magnetic string has a nonzero angular component \cite{Aharonov59}
\begin{equation}\label{abp}
eA_{\varphi} = {e\Phi \over 2\pi\rho} = -{\Phi \over \Phi_{0}\rho} =
{\phi\over\rho},
\end{equation}
where $\Phi$ is the magnetic flux and $\Phi_{0}=2\pi \hbar c/|e|$ is the
magnetic flux quantum. It describes the singular magnetic field with support on the $z$ - axis
\begin{equation}\label{B}
B_z = {\phi\over e\rho}\,\delta (\rho).
\end{equation}
It is the fractional part $\delta$ of the magnetic flux $\phi = N+\delta,\; 0<\delta<1$ which produces the physical effects. Its integer part $N$ appears as a phase factor $\exp(iN\varphi)$ in solutions of the Dirac equation, therefore it does not influence physical quantities and can be dropped. 

Written in the cylindrical coordinates $\rho,\;\varphi,\;z,$ the Dirac operator
\begin{equation} \label{de}
H = \alpha_{i}(p_{i} - eA_{i}) + \beta M
\end{equation}
together with the operators 
\begin{equation} \label{co}
\hat{p_3} = -i\partial_z, \quad \hat{J_3} = (-i\partial_{\varphi}+{1 \over 2} \Sigma_3), \quad \hat{S_3}=\beta\Sigma_3+\gamma \;{p_3\over M}, 
\end{equation}
($\alpha, \beta, \gamma$ is the known matrices), form a complete set of commuting operators. The eigenfunctions of this set are the cylindrical modes marked by the quantum numbers of the energy $E=\sqrt{p^2+M^2}=\sqrt{p_{\perp}^2+p_3^2+M^2},$ of the $z$ - components of linear and total angular momenta $p_3$ and $j_3 = l+{1\over2},$ respectively, and by the quantum number $s=s_3\sqrt{1+p_3^2/M^2},\,s_3=\pm 1,$ which describes the electron spin polarization in the magnetic field  (in nonrelativistic limit it is $z$ - component of spin). Instead of the half-integer quantum number $j_3$ we introduce the integer number $l$.

The radial functions of these cylindrical modes are the Bessel functions of the first kind of positive and negative order with arbitrary coefficients. The normalization condition for the partial modes with quantum numbers $j = (p_{\perp},p_3,l,s),$
\begin{equation} \label{nc}
\int dx \psi^{\dagger}(j,x) \psi(j^{\prime},x) = \delta_{j, j^{
\prime}} = \delta_{s, s^{ \prime}}\,\delta_{l, l^{ \prime}}\,
\delta(p_{3} - p_{3^{ \prime}})\,{\delta(p_{ \perp} - p_{ \perp}^{
\prime}) \over  \sqrt{p_{\perp} p_{ \perp}^{ \prime}}} \,,
\end{equation}
fixes the solutions (electron states with $E>0$) for values of $l\neq 0.$ They  contain only the regular Bessel functions of positive order. The solution for $l=0,$ on the other hand, is the sum of regular and irregular but square intergrable Bessel functions of the orders $-\delta$ or $-1+\delta$.  It is impossible to remove the irregular components from this mode by any choice of the coefficients. Obviously this fact is the consequence of an additional attraction caused by the interaction between spin and magnetic field. It is connected with the general problem of the self-adjoint extension for the Hamilton operator in the presence of the singular AB potential. The correct choice of the mode with $l=0$ can be obtained by a limiting procedure \cite{Hagen90} based on more realistic models of the magnetic field with regular potentials.

Taking into account the orientation of the magnetic field (\ref{B})
we obtain the total mode functions
\begin{eqnarray}\label{es}
\psi_{e}(j,x) = {1\over 2\pi}{1\over\sqrt{4sE_p}}\;
e^{-iE_p t + ip_3z}\, e^{i{\pi\over 2}|l|}\,
\left(\begin{array}{c} \displaystyle \sqrt{E_p +
sM}\sqrt{s+1}\;J_{\nu_1}(p_{\perp}\rho)\, e^{il\varphi}\\
\displaystyle i\epsilon_3\epsilon_{l}\sqrt{E_p - sM}\sqrt{s-1}\;
J_{\nu_2}(p_{\perp}\rho)\, e^{i(l +1)\varphi} \\
\displaystyle \epsilon_3 \sqrt{E_p + sM}\sqrt{s-1}\;J_{\nu_1}(p_{\perp}\rho)\,
e^{il\varphi}\\ \displaystyle i\epsilon_{l}\sqrt{E_p - sM}\sqrt{s+1}
\; J_{\nu_2}(p_{\perp}\rho)\, e^{i(l +1)\varphi}
\end{array}\right)
\end{eqnarray}
where
\begin{equation} \label{order1}
\nu_1 := \left\{\begin{array}{r} l-\delta \\ -l+\delta
\end{array}\right., \; \nu_2 := \left\{\begin{array}{r}
l+1-\delta \\ -l-1+\delta \end{array}\right., \; \epsilon_l :=
\left\{\begin{array}{rl} 1 &\quad {\rm if}\ \ l \geq 0 \\ -1& \quad
{\rm if}\ \ l<0 \end{array}\right., \; \epsilon_3 := {\rm sign}(s p_3)\,.
\end{equation}

The complete set of solutions to the Dirac equations includes the electron states of negative energy. Instead of using them we introduce positron states $\psi_{p}$ with $E>0$ which can be obtained from electron states of negative energy by the charge conjugation operation
\begin{equation}
\psi \rightarrow \psi_{c} = C\bar \psi_{transp} \,, \quad C = \alpha_{2}.
\end{equation}
Since the function $\psi_{c}$ has the quantum numbers $E, -p_3, -j_3, s$ one needs to replace $p_3 \rightarrow -p_3,\;j_3 \rightarrow -j_3 \;(l \rightarrow -l-1)$ in the electron state of negative energy to obtain from $\psi_{c}$ the positron state $\psi_{p}$ with quantum numbers $E, p_3, j_3, s.$

The cylindrical modes obtained above are the exact solutions of the Dirac equation in the AB potential. But these quantum states do not describe ingoing and outgoing particles with definite linear momenta at infinity. In order to calculate the cross sections of the QED processes in terms of particle fluxes at infinity we need the electron and positron {\em scattering wave functions}. There exist two independent exact solutions of the Dirac equation
$\Psi^{(-)}(J,x)$ and $\Psi^{(+)}(J,x)$ which behave at large distances like a plane wave propagating in the direction $\vec{p}$ (given by $p_x = p_{\perp} \cos\varphi_p, \; p_y = p_{\perp} \sin\varphi_p, \; p_z=p_3)$ plus ingoing or outgoing cylindrical waves, correspondingly. They can be obtained as superpositions of the cylindrical modes. To describe the interaction of the charged particle with the external field correctly \cite{Akhiezer65}, one needs to take $\Psi^{(+)}(J,x)$ for ingoing particles and $\Psi^{(-)}(J,x)$ for outgoing ones. $J$ is a collective index for $\vec{p}$ and $s$. 

\bigskip

To complete the description of the involved particle fields we notice that the external AB field does not influence on the photon wave function. In cylindrical coordinates it reads
\begin{equation}\label{vp}
A_{\mu}^{\lambda}(\vec{k}, x) =
{e_{\mu}^{(\lambda)}\over\sqrt{2\omega_k}}e^{-i\omega_k t + ik_3 z}
e^{ik_{\perp}\rho \cos(\varphi-\varphi_k)}
\end{equation}
where the polarization vectors
\begin{equation}\label{polv}
e^{(\sigma)} := (0,\; -\sin \varphi_k, \;\cos \varphi_k, \; 0) \,, \quad
e^{(\pi)} := {1\over \omega_k}(0,\; -k_3\cos \varphi_k,\; -k_3\sin
\varphi_k,\; k_{\perp})
\end{equation}
correspond to two linear transversal polarization states. In the
coordinate frame with $k_3=0$ the polarization vector $e^{(\pi)}$ is
directed along $z$-axes and $e^{(\sigma)}$ is normal to the magnetic
string.


\section{Matrix elements for the bremsstrahlung \newline
and pair production processes}

Differential cross sections of the processes are related to plane wave states or, in external fields, to the scattering states. The cylindrical modes have a vanishing radial flux and therefore do not describe ingoing or outgoing particles. They are, however, convenient for calculating matrix elements, and we use these matrix elements as starting point for calculation of the differential cross section which refers to scattering states.

\bigskip

The matrix element for the bremsstrahlung process for the transition of an ingoing electron with quantum numbers $j_p = (p_{\perp}, p_3,l,s)$ to an outgoing electron with quantum numbers $j_q = (q_{\perp}, q_3,n,r)$ and a photon with quantum numbers
($\vec{k},\; \lambda)$ has the usual form
\begin{equation}\label{brme}
\widetilde{M}_{\lambda}(j_q, \vec{k},\lambda; j_p) = -i\,\langle j_q, \vec{k},\lambda|S^{(1)}|j_p \rangle = - e \int {d^4x}\bar\psi_{e}(j_{q},x)
\;{A^{\ast}}^{\lambda}_{\mu}(\vec{k},x)\gamma^{\mu}\;\psi_{e}(j_{p},x).
\end{equation}
\noindent
The matrix element for the pair production of an electron with quantum numbers $j_p = (p_{\perp}, p_3,l,s)$ and a positron with quantum numbers $j_q = (q_{\perp}, q_3,n,r)$ by a single photon with quantum numbers ($\vec{k},\; \lambda)$ for physical states $\lambda = \sigma,\; \pi$ reads correspondingly
\begin{equation} \label{ppme}
\widetilde{M}_{\lambda}(j_p, j_q; \vec{k}, \lambda) = -i\,\langle j_q, j_p|S^{(1)}|\vec{k}, \lambda \rangle = -e \int {d^4x}\bar\psi_{e}(j_{p},x)
\;A^{\lambda}_{\mu}(\vec{k},x)\gamma_{\mu}\; \psi_{p}^c (j_{q},x),. \end{equation}

These matrix elements were calculated in \cite{Audretsch96'} and \cite{Audretsch96''}. We will point out here only some specific results of the calculation. The matrix elements contain two $\delta$-functions which correspond to the energy and $z$-component of linear momentum conservation laws. $J_3$-component of total angular momentum is conserved too. On the contrary, the linear momenta in the plane perpendicular to the magnetic string are not conserved. Moreover, the processes can happen only if the radial
momentum of an ingoing particle is bigger than the sum of the radial momenta of the outgoing particles. This is in the accordance with the energy conservation law. An excess of the radial momentum is transmitted to the flux tube.

This result for the matrix elements follows from the integral factors of the type (see formulae 6.578(3), 6.522(14) of
\cite{Gradshteyn80})
\begin{eqnarray} \label{int}
J(\alpha, \beta) &:=& \int_{0}^{ \infty} \rho d\rho
J_{\alpha}(x\rho\sin{A}\cos{B})
J_{\beta}(x\rho\cos{A}\sin{B})
J_{\beta-\alpha}(x\rho) = \nonumber\\
&=& {2\sin\pi\alpha\over \pi x^2
\cos(A+B)\cos(A-B)}\left({\sin{A} \over \cos{B}}
\right)^{\alpha}\left({\sin{B} \over \cos{A}}
\right)^{\beta}
\end{eqnarray}
which arise after integration over $\varphi$ and $\rho$.
Here $x,\, x\sin{A}\cos{B},\, x\sin{B}\cos{A}$ are the radial momenta of the ingoing particle and the outgoing particles, respectively. $\alpha$ and $\beta$
are the Bessel indices of the radial modes.

The partial wave analysis of the processes shows a rather unexpected
feature: The processes turn out to be forbidden unless the quantum
numbers $l$ and $n$ of the charged particles have opposite signs.
This in turn implies that the expectation values of their kinetic
angular momentum projections, $<[\vec{r}\times (\vec{p}-e\vec{A})]_3>=<- i \partial_\varphi -\phi>$, have opposite signs for all values of  $l,\;n \neq0$.
In the framework of a semiclassical picture this leads to the interpretation that the charged particles pass the magnetic string in opposite directions, encircling the magnetic flux. Only then both the AB--effect itself and the accompanying effects can happen. Apparently the constellation is necessary for the ingoing particle to give the excess of its radial momentum to the string and create real outgoing particles from vacuum fluctuations. We notice also that the specific AB-factor $\sin \pi\delta$ arises in the matrix elements.

\bigskip

It is now possible to calculate the matrix elements taken with respect to the
scattering states in terms of the matrix elements (\ref{brme}), (\ref{ppme}) for the cylindrical modes and to find the cross sections of the processes.

For the bremsstrahlung process of an electron with momentum $\vec{p}$ and spin $s$ which is scattered by the magnetic string and emits a photon with momentum $\vec{k}$ and polarization $\lambda$ we have
\begin{equation}
M_{\lambda} := -i\,\langle \vec{q}, r; \vec{k}, \lambda) |S^{(1)}|\vec{p}, s\rangle = \sum_{l, n} c^{(+\,e)}_l {c_{n}^{(-\,e)}}^{\ast}
\widetilde{M}_{\lambda}(j_q, \vec{k},\lambda; j_p)
\end{equation}
where the coefficients are equal to 
\begin{equation}
c^{(\pm\,e)}_l := e^{-il\varphi_p}\;e^{\pm i{\pi\over 2}\epsilon_l
\delta}
\end{equation}
and $\vec{q}$ and spin $r$ are the quantum number of the scattered electron.

For production of an electron-positron pair with momenta
$\vec{p},\;\vec{q}$ and spins $s,\;r$, correspondingly by an
incoming photon with momentum $\vec{k}$ and polarization $\lambda$
we find
\begin{equation}
M_{\lambda} := -i\,\langle \vec{q}, r; \vec{p}, s)|S^{(1)}|\vec{k},
\lambda)\,\rangle = \sum_{l, n} c_l^{(-\,e)\ast} c_n^{(-\,p)}\;
\widetilde{M}_{\lambda}(j_p, j_q; \vec{k}, \lambda). 
\end{equation}
with the coefficients 
\begin{equation}
c^{(-\,p)}_n := e^{i(n+1)(\varphi_q+\pi)}\;e^{i{\pi\over
2}\epsilon_n \delta}.
\end{equation}

With these matrix elements one can calculate the effective differential cross sections for the bremsstrahlung \cite{Audretsch96'} and the pair production  \cite{Audretsch96''} processes. They contain the complete information about the energy, momentum and polarization distributions of the created particles. 
The corresponding exact results are rather cumbersome to be presented here. They simplify considerably for low energies of the incoming particles. We will show below that the low energy regime is more suitable for possible experiments with magnetic strings.

\section{The cross sections at low energies}

In this section we will analyze the differential and total cross sections for bremsstrahlung and pair production processes in the Aharonov--Bohm potential at low energies of the incoming particles.

\subsection{Bremsstrahlung}

For low electron energy, $v\ll 1$, the differential cross section per unit length of the magnetic string takes the form \cite{Audretsch96'}
\begin{equation}  \label{brdcsle}
{d\sigma_{\lambda}\over d\omega_k d\Omega_k d\varphi_q} =
{e^2\;\sin^2\pi\delta \over 32\pi^4\;M\;\omega_k}\;v\;\Theta(sr)\;S^{(s)}
\pmatrix{1 \cr
\cos^2\vartheta_k \cr} \quad {\rm for}\quad
\cases{ \lambda=\sigma\cr
          \lambda=\pi \cr} 
\end{equation}
with
$$
S^{(s)}=\left(1-{\omega_k \over E_p-M}\right)^{1+s\delta} +
\left(1-{\omega_k\over E_p-M}\right)^{-s\delta} \pm 2
\left(1-{\omega_k\over E_p-M}\right)^{1\over2}
\cos(\varphi_{pk}+\varphi_{qk})\,,
$$
where the sign $\pm$ stands for $\lambda=\sigma,\,\pi.$ The cross section of the bremsstrahlung process is proportional to the classical electron radius, $r_0 = e^2/4\pi M,$ and the velocity $v$ of the ingoing electron.

Let us note some particular features of the bremsstrahlung process for the Dirac electron which can be read of from eq. (\ref{brdcsle}).
\begin{itemize}
\item[(i)]
The electron spin projection is conserved at low electron energies.
\item[(ii)]
The appearance of the additional factor $\cos^2\vartheta_k$ for the photon $\pi$ - polarization state is typical for the angular distribution of radiation from non-relativistic particles. But the angular distributions of the outgoing electron and emitted photon are not independent on the angles $\varphi_q$ and $\varphi_k$ as it is the case for spinless particles \cite{Sereb88}. Apparently the angular asymmetry of the process in the plane perpendicular to the magnetic string arises due to the interaction of the electron's magnetic moment with the
magnetic field. After integration over the angle $\varphi_q$ of the outgoing electron this asymmetry disappears.
\end{itemize}

Integrating over final states, we find the total cross section at low electron energy
\begin{equation} \label{tcsle}
{d{\sigma}_{\sigma}\over d\omega_k} = 3 {d{\sigma}_{\pi}\over
d\omega_k} = {r_0\;\sin^2\pi\delta \over \pi} {v\over\omega_k}
\left[\left(1-{\omega_k \over E_p-M}\right)^{1+s\delta} +
\left(1-{\omega_k\over E_p-M}\right)^{-s\delta}\right]\,.
\end{equation}
The cross section coincides for $s=-1$ with the cross section for non-relativistic spinless particles \cite{Sereb88}. For $s=1,$ however, the interaction between the spin and the magnetic field is attractive, the wave function is amplified near the string, and this results in a modification of the cross section.

\subsection{Pair production}

One can show that the electron--positron pairs of low energies are created
mainly from $\pi$ - polarized photons. Above the threshold $\omega_k - 2M \ll M$ the differential cross section per unit length of the magnetic string is in this case \cite{Audretsch96''}
\begin{equation} \label{ppdcsle}
{d\sigma_{\pi}\over d E_q d\varphi_q d\varphi_p dq_3} \approx
{e^2\;\sin^2\pi\delta\over 256\pi^4 } (1-s_3 r_3){c^{-\delta}
(1+s_3) + c^{\delta} (1-s_3)\over M^3}
\end{equation}
with $c \approx p_\perp^2 q_\perp^2/16 M^4 \ll 1.$
\noindent
The angular distributions for electrons and positrons of low energies are uniform in the plane perpendicular to the magnetic string but their dependence on the polar angle $\vartheta $ is rather intricate.

The polarizations of the electron and the positron depend strongly on the photon polarization state. For $\pi$-polarized photons the created particles have spin projections of the opposite signs, and electrons with positive spin projections (antiparallel to the magnetic string) are produced predominantly since $c\ll 1.$  In this case the interaction of the magnetic moments with the string magnetic field is attractive for both the electron and the positron, and their wave functions are localized near the string. The $\sigma$-polarized photon, on the other hand possesses a polarization vector perpendicular to the magnetic string and creates particles with the spin projections $s_3$ and $r_3$ of equal signs which implies that their magnetic moments have opposite directions. In this case, because of the interaction of the magnetic moments with the string magnetic field one of the particles is localized near the string while the other one is localized at a certain distance, and the pair production process is suppressed.

Performing the integration over final states we obtain the total cross section of pair production at photon energies above the pair production threshold \begin{equation}\label{pptcsle}
\sigma_{\pi} \sim {r_0\;\sin^2\pi\delta\over \sqrt{2}\pi}\;\left({\omega_k-2M
\over 2M}\right)^{{3\over 2}-2\delta}.
\end{equation}
After integration over a small energy interval $\Delta$ above the threshold, $2M \ge \omega_k \le 2M (1+\Delta)$ the integral cross section for pair production reads
\begin{equation} \label{I}
I := \int_{2M}^{2M(1+\Delta)}\sigma_{\pi}(\omega_k)\,d\omega_k \sim
{e^2\,\sin^2\pi\delta\over 2\sqrt{2}\pi}\;{\Delta^{{5\over 2}-2\delta} \over {5\over 2}-2\delta}.
\end{equation}
This quantity determines the output of electron--positron pairs produced by a photon per unit time and per unit length of the magnetic string within the given energy interval.


\section{Conclusions}

We have analyzed the QED processes, bremsstrahlung of a scattered electron and electron-positron pair production by a single photon under the influence of a magnetic string. We presented the differential cross sections of the processes, which contain complete information about energy, angular and polarization distributions of the created particles, as well as the total cross section and analysed them at low energies. In addition to the AB influence which all quantum particles suffer, spin particles interact with the magnetic field via their magnetic momenta. This interaction influences strongly the behavior of wave functions near the flux tube and leads to a specific behavior of the cross sections. For the idealized case of an infinitely thin magnetic string the wave functions do not vanish on the string, and, in a way, the nonlocality of the AB effect is lost.

Let us add a few remarks regarding the experimental verification of the results obtained. It is necessary to notice that  the AB regime can be realized only for particle wave lengths bigger than diametrical size of the string.
We found that the differential cross sections of the QED processes in the presence of the AB magnetic string turned out to be not extremely small even for low energy particles and small magnetic fluxes. This surprising conclusion contrasts with the well-known results for the QED processes in an uniform magnetic field where the quantum effects manifest themselves only for relativistic particles or in intense magnetic fields \cite{Klepikov54,Sokolov68,Ritus79}. This happens at energy $E$ and field strength $B$ for which the typical parameter \cite{Schwinger51} is of order 1, 
\begin{equation}\label{Schwinger}
{B\over B_0}{E\over Mc^2} \sim 1,\quad {\rm where} \quad B_0 = {M^2c^3\over e\hbar}\,.
\end{equation}
At present these energies and magnetic fields are beyond the experimental possibilities. 

That there is such a difference in the AB situation can be understood in following way: The string magnetic field is singular and therefore the criterion (\ref{Schwinger}) is formally valid at any energy. Certainly, the magnetic string is a limiting case of a realistic model of a thin solenoid which contains an intense but finite magnetic field. The results of papers \cite{Audretsch96',Audretsch96''} sketched above are valid only in this limiting case, and it would therefore be very important to know how they change if the finite radius of the solenoid is taking into account. We are working on this problem.

In any case, already the observation of the ordinary AB effect, which is done by means of electron interference and electron holography \cite{Peshkin89}, is not a simple task. This is even more the case for QED processes in this context.  The AB bremsstrahlung effect with polarized electron beams requires a careful discussion and preparation. The same is true for the AB pair production from single photons. For these photons of rather high energies there exist additional effects which can obscure the AB--pair production. In particular, this is the pair production by photons in collision with material of the tube carrying the magnetic flux. 

\bigskip

The work was supported by the Deutsche Forschungsgemeinschaft and by Russian Foundation for Basic Research (96-02-16053-a).


\end{document}